\documentclass[aps,prl,reprint,groupedaddress,showpacs]{revtex4-1}
\usepackage{hyperref}
\usepackage{graphicx}
\usepackage{amsmath}

\begin{document}


\title{Symmetry Protected Topological Phases and Majorana Mode in One-dimensional Quantum Walk with Boundary}


\author{Ho Tat Lam}
\author{Yue Yu}
\author{Kwok Yip Szeto}
\affiliation{Department of Physics, Hong Kong University of Science and Technology, Hong Kong}
\date{\today}



\begin{abstract}
The topological phases in one-dimensional quantum walk can be classified by the coin parameters. By solving for the general exact solutions of bound states in one-dimensional quantum walk with boundaries specified by different coin parameters, we show that these bound states are Majorana modes with quasi-energy $E=0,\pi$. These modes are qualitatively different for different boundary conditions used. For two-boundary system with symmetric boundary conditions, the interaction energy between two Majorana bound states can be computed, as in the case of a finite wire. Suggestion of observing these modes are provided. 
\end{abstract}


\maketitle


- \textit{Introduction} Quantum walk provides an excellent platform for us to further our understanding of the interference and superposition features that lead to a non-classical dynamic evolution. The recent advances in discrete time quantum walk \cite{1,2,3,4,5} have created not only many interesting research endeavors in its application, such as in quantum computation \cite{6,7,8,9,10,11}, but also in the possibility of finding new topological phases in condensed matter physics \cite{12}. In discrete quantum walk \cite{3,4,13,14,15,16,17,18,19}, the many parameters characterizing the quantum coin allow the theorists to explore different physical scenario for the walker, thereby relating theoretical prediction with experiments for the studies of different dynamic scenarios \cite{20,21}. 

For system with two internal degrees of freedom, the quantum coin is a 2$\times$2 unitary matrix, where Hadamard coin is usually employed. Here we consider a generalization of the Hadamard coin with a rotational matrix parametrized by a parameter $\theta\in(-\pi,\pi)$,
\begin{equation}
C=\left[ \begin{array}{cc}
\cos{\theta} & \sin{\theta} \\
-\sin{\theta} & \cos{\theta} \end{array} \right]\label{eq:1}
\end{equation}
The basis for this matrix are $\{|L\rangle,|R\rangle\}$, which are the two internal degree of freedom of the walker, with the left (L) and right (R) labels for one-dimensional quantum walk. The quantum walk is implemented by an iteration of a unitary transformation acting on a two-state particle, with a wave function $|\psi\rangle=\sum\left(a_n|L\rangle+b_n|R\rangle\right)|n\rangle$ 
defined on a discrete one-dimensional space where $|n\rangle$ is the state that quantum walk localized at position of integer $n$, with 
$a_n$ and $b_n$ being the left and right components of the wave function at position $n$. 
The unitary transformation $U = SC$ consists of a quantum coin operation $C$ that acts on the internal state, followed by a shift operator $S$ to evolve the particle coherently in position space. 
The shift operator $S=|L\rangle{\langle}L|\sum|n\rangle{\langle}n+1|+|R\rangle{\langle}R|\sum|n\rangle{\langle}n-1|$ 
moves the left and right component to the left and right direction by one unit respectively. 
The eigenstates of a quantum walk which coin is translational invariant are plane waves characterized by their quasi-momentum $k$, with quasi-energy $E$ and is written as  
$\psi_k(n,t) = e^{-iEt+ikn}  \left[a_k, b_k\right]^T$.  
The quasi-momentum and quasi-energy have a periodicity of $2\pi$ and we restrict the value to be from $(-\pi,\pi)$ (from now on, we omit the prefix "quasi-" in front of energy and momentum). From the evolution equation, the energy $E$ and the momentum $k$ satisfies the dispersion relation $\cos E=\cos\theta\cos k$ and the normalized vector $\left[a_k, b_k\right]^T$ is,
\begin{equation} 
\frac{1}{\sqrt{2\sin E(\sin E+\cos\theta\sin k)}}\left[\begin{array}{c}
i\sin\theta e^{ik}\\
\sin E+\cos\theta\sin k\end{array} \right]\label{eq:2}
\end{equation}
The dispersion relation shows that for a given momentum $k$, there exist two eigenstates with energy $E=\pm\arccos(\cos\theta\cos k)$, which we denote them by  $|k,+\rangle$ and $|k,-\rangle$ , corresponding to the positive and negative energy eigenstates respectively. All the eigenstates form two energy bands with a gap which closes at $E=0$ and $E=\pi$ when $\theta=0$ or $\theta=\pi$. 

- \textit{Topological Phases} Recently T. Kitagawa has shown that quantum walks have a nontrivial winding number of 1 due to its hidden chiral symmetry and for the split-step quantum walks there can be two topological phases \cite{22}. 
Firstly, we write the unitary transformation $U(k)$ for a specific momentum  $k$ as
\begin{equation}
\begin{aligned}
&U(k)=\left[\begin{array}{cc}
\cos\theta e^{ik} & \sin\theta e^{ik}\\
-\sin\theta e^{-ik} & \cos\theta e^{-ik}
\end{array}\right]\\
\end{aligned}\label{eq:3}
\end{equation}
which can be rewritten in the form of the evolution operator 
$U(k)=e^{-iH(k)}$ provided that $H(k)=E(k)\left[\textbf{n(k)}\cdot\vec{\sigma}\right]$ with  $E(k)$ chosen to be positive and the unit vector $\textbf{n(k)}$ defined by,
\begin{equation}
\textbf{n(k)}=-(\sin\theta\sin k,\sin\theta\cos k,\cos\theta\sin k)/\sin E \label{eq:4}
\end{equation}
One can verify that 
$U(k)=(\cos E)\vec{I}-i\sin E\left[\textbf{n(k)}\cdot\vec{\sigma}\right]$. 
Here $\vec{I}$ is the identity matrix and $\vec{\sigma}$ the Pauli matrix. The Hamiltonian $H(k)$ can also be written in terms of the eigenstates, $H(k)=E|k,+\rangle\langle k,+|-E|k,-\rangle\langle k,-|$. This Hamiltonian $H(k)$ can be mapped to a unit sphere through the vector $\textbf{n}(k)$ that traces out a circle on the equator of the unit sphere when $k$ goes from $-\pi$ to $\pi$ (see Fig. 1) and the circle provides a winding number corresponding to the topological number of the quantum walk. 
\begin{figure}
\includegraphics[scale = 0.28]{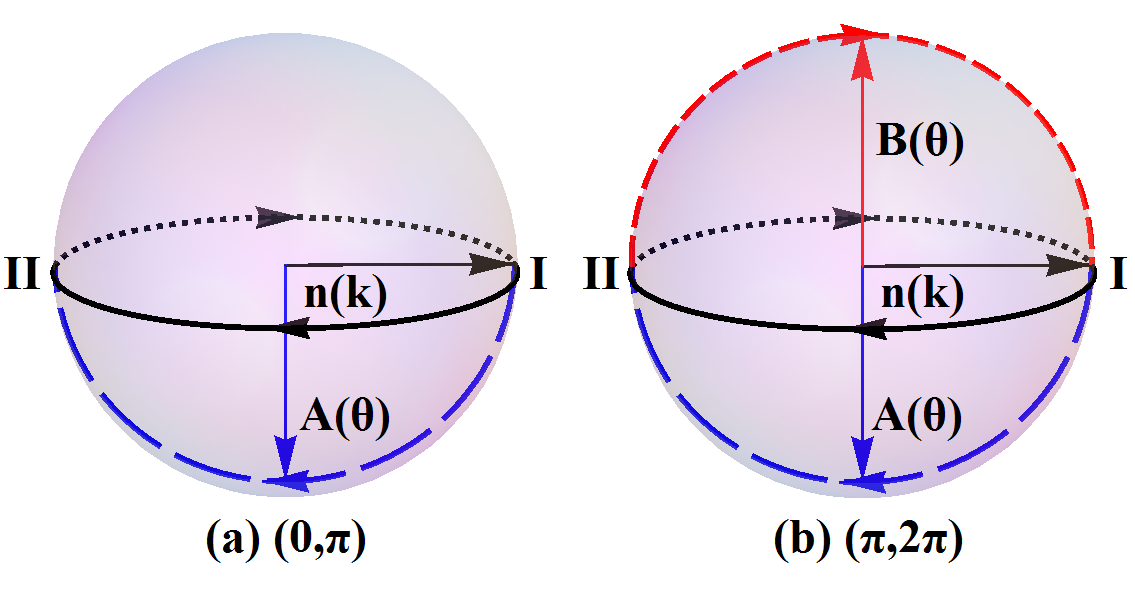}
\caption{\label{Figure 1}The trajectory of vectors $\textbf{n}(k)$ (black circle) and $\textbf{A}(\theta)$ on the unit sphere (blue semi-circle). The vector $\textbf{A}(\theta)$ is perpendicular to the black curves and defines the winding direction of the vector $\textbf{n}(k)$}
\end{figure}
The sense of the winding of the circle around the equator can be defined through the normal vector $\textbf{A}(k)$ such that $\textbf{A}\cdot\textbf{n}(k)=0$,  
\begin{equation}
\textbf{A}(\theta)=\frac{\textbf{n}(0)\times\textbf{n}(\pi/2)}{|\textbf{n}(0)\times\textbf{n}(\pi/2)|}=\mathrm{sgn}(\sin\theta)(\cos\theta,0,-\sin\theta)
\end{equation}
As k changes, $\textbf{n}(k)$ moves on a plane with normal $\textbf{A}(k)$. This is due to the hidden chiral symmetry in the  Hamiltonian: 
\begin{equation}
\Pi^{-1} H(k)\Pi=-H(k), \quad\Pi=e^{iA\cdot\sigma \pi/2}
\end{equation}
where $\Pi$ is the chiral symmetry operator which rotates $\textbf{n}(k)$ by $\pi$ around the axis defined by $\textbf{A}(k)$. This chiral symmetry is essential for the existence of topological phase in one-dimension and the resultant topological phases are called symmetry protected topological phases. With this chiral symmetry, the Hamiltonian can be off-diagonalized and the topological number is determined by the off-diagonal terms \cite{24}, which here is achieved by the application of a unitary transformation $\Lambda=e^{-i\sigma_y(\pi/2-\theta)/2}$, 
\begin{equation}
\Lambda^{-1} H \Lambda=\frac{E}{\sin E}\left[\begin{array}{cc}
0&h(k)\\
h^{\dagger}(k)&0
\end{array}\right], h=\sin k-i(\sin \theta\cos k)
\end{equation}
The unitary transformation $\Lambda$ is a rotation around y-axis (which is along $\textbf{n}(0)$) by $\pi/2-\theta$. It transforms $H(k)$ to a new Hamiltonian that has only $\sigma_x$ and $\sigma_y$ components, so that as $k$ changes, it traces out a circle on a 2-dimensional $x-y$ plane. The topological number can then be defined as the winding number of $h$ around the origin,
\begin{equation}
m=\frac{1}{2\pi i}\int^{\pi}_{-\pi}\mathrm{d} k \frac{d}{dk}\ln h(k)=\mathrm{sgn}(\sin\theta)
\end{equation}
The discontinuity of topological number $m$ at $\theta=0,\pi$ can be understood as the discontinuity of the vector $\textbf{A}(\theta)$. As $\theta$ increases from $0$ to $\pi$, $\textbf{A}(\theta)$ goes from point I to point II through the blue dash semicircle shown in Fig. 1. In this regime, $\textbf{A}(\theta)=(\cos\theta,0,-\sin\theta)$. 
As $\theta$ increases from $\pi$ to $2\pi$, (or equivalently from $-\pi$ to $0$) 
$\textbf{A}(\theta)=(-\cos\theta,0,\sin\theta)$ and we see that it jumps from point II to point I and goes through the blue dash semicircle once again. 
To handle this discontinuity, we introduce another continuous vector  $\textbf{B}(\theta)=(\cos\theta,0,-\sin\theta)$, which does not contain the sign function 
$\mathrm{sgn}(\sin\theta)$. 
For $\theta$ between $0$ and $\pi$, $\textbf{B}(\theta)=\textbf{A}(\theta)$; while for $\theta$ between $\pi$ and $2\pi$, 
$\textbf{B}(\theta)=-\textbf{A}(\theta)$. 
The positive winding orientation is defined to be along vector $\textbf{B}(\theta)=(\cos\theta,0,-\sin\theta)$ which goes through a complete circle continuously from point I to II and then from II to I again. Accordingly, the topological numbers should be $m=\vec{A}\cdot \vec{B}=\mathrm{sgn}(\sin\theta)$. The ambiguity in the definition of topological number occurs only at $\theta=0,\pi$ because the band gap closes at these points. The absolute value of the topological number is the number of times that the loop winds around the equator, which gives a winding number $|m|=1$, but there exist two topological phases with topological number $m=\pm1$. 

- \textit{Majorana Boundary Mode} From bulk-boundary correspondence \cite{23}, there should be two gapless states localized at the surface when these two topologically distinct phases are joined together. These gapless states occurs only at $E=0,\pi$, thus there should be two bound states of energy either at $E=0$ or $E=\pi$ on the boundary of quantum walk with $\sin\theta$ of different sign. Based on these observations, we seek to find these boundary states by considering a boundary at the origin ($n=0$) where the quantum walk has different coins on the two sides of the boundary. Specifically, $\theta_1$ is used for $n\leq0$ and $\theta_2\ne \theta_1$ for $n\geq1$. Following the analysis of Ramsauer effect in quantum walk\cite{25}, we look for bound states for this quantum walk by the following ansatz, with the condition that the wave function decays to zero at infinity,
\begin{equation}
\begin{aligned}
\psi(n)&=re^{\kappa_1 n}\left[\begin{array}{c}
a_{-i\kappa_1}\\
b_{-i\kappa_1}
\end{array}
\right]&n\leq0\\
&=te^{-\kappa_2 n}\left[\begin{array}{c}
a_{i\kappa_2}\\
b_{i\kappa_2}
\end{array}
\right]&n\geq1
\end{aligned}\label{eq:11}
\end{equation}
Here we replace the real $k$ in $\left[a_k, b_k\right]^T$ with an imaginary one. For $n\leq0$, the spinor is determined by $\theta_1$; for $n\geq1$, the spinor is determined by $\theta_2$. Using the dispersion relation $\cos E=\cos\theta\cos k$, we can relate the energy of this hypothetical bound state with the decay constant $\kappa_1$ and $\kappa_2$,
\begin{equation}
\cos E=\cos\theta_1\cosh\kappa_1=\cos\theta_2\cosh\kappa_2\label{eq:12}
\end{equation}
By imposing the continuity conditions on the bound state wave function at $n=0,1$, we get the relations between the $r$ and $t$:
\begin{equation}
\begin{aligned}
ra_{-i\kappa_1}&=ta_{i\kappa_2},& re^{\kappa_1}b_{-i\kappa_1}&=te^{-\kappa_2}b_{i\kappa_2}\label{eq:13}
\end{aligned}
\end{equation}
where the first equation comes from left component of $\psi(0)$ and the second comes from the right component of $\psi(1)$. In order for the bound state to exist, we need the existence of nontrivial solution of $r$ and $t$, which needs $\sin\theta_1$ and $\sin\theta_2$ to be of opposite sign (refers to Appendix I in the Supplementary Material). This implies that two sides of the boundary are topologically distinct. Under this condition, there are always two bound states solution with energy $E=0,\pi$, which comes in pair. This is also predicted from the bulk-boundary correspondence.

For a concrete example of quantum walk where the coin parameter is positive on the left and negative on the right of the origin, the bound states solutions are,
\begin{equation}
\begin{aligned}
E=0\quad\psi(n)&=x_1^{n-1}\left[\begin{array}{c}
x_1\\
-1
\end{array}
\right]&n\leq0\\
&=x_2^{n-1}\left[\begin{array}{c}
x_2\\
-1
\end{array}
\right]&n\geq1\\
E=\pi\quad\psi(n)&=(-x_1)^{n-1}\left[\begin{array}{c}
x_1\\
1
\end{array}
\right]&n\leq0\\
&=(-x_2)^{n-1}\left[\begin{array}{c}
x_2\\
1
\end{array}
\right]&n\geq1\\
\end{aligned}
\end{equation}
where $x_1$ is $(1+\sin\theta_1)/\cos\theta_1$ and $x_2$ is $(1+\sin\theta_2)/\cos\theta_2$. 
The two sides of the wave function only depend on the coin parameter of their own sides. For two regions with same topological number, the wave function on both regions can be joined together and the continuity condition on the boundary will be satisfied automatically. This observation is important in our discussion for finite wire.  

Now that our ansatz for the boundary state is possible, we claim that it is a Majorana bound state, which is a $0+1$ dimensional particle whose antiparticle is itself \cite{26}, ( $0+1$ stands for $0$ dimension in space and $1$ dimension in time). The system has an intrinsic particle-hole symmetry. We denote the complex-conjugation operator as $K$. Then, the eigenstate with momentum $k$ and energy $E$ and the eigenstate with momentum $-k$ and energy $-E$ are related by an antiunitary operator $\Gamma=-K$, $|k,E\rangle=\Gamma |-k,-E\rangle$. This antiunitary operator is the particle-hole operator so that our effective Hamiltonian has an intrinsic particle-hole symmetry,
\begin{equation}
\Gamma H(k) \Gamma^{-1}=(-K)H(k)(-K)=-H(-k)
\end{equation}
This particle-hole symmetry shows that a creation of a particle with energy $E$ is equivalent to the annihilation of a hole with energy $-E$. Because of the periodicity of the energy, $E$ and $-E$ coincide when $E=0,\pi$, the creation and annihilation operator are the same for $E=0$ and $E=\pi$ and the boundary modes are Majorana modes.

- \textit{Finite Wire} We now consider quantum walk on a finite wire. Note that when $\theta_1=\pm\pi/2$, the decay constant $\kappa_1$ in the bulk is infinity according to the dispersion relation in Eq. \ref{eq:12} and thus the quantum walk does not penetrate through a site with coin parameter $\theta_1=\pm\pi/2$ which corresponds to 
\begin{equation}
C=\pm\left[\begin{array}{cc}
0&1\\
-1&0
\end{array}\right]
\end{equation}
This coin interchanges the left and right components with or without an extra sign so that it reflects the quantum walk. Thus, the proper boundary conditions describing the finite wire is to set the coin parameters at the two ends of the wire to be $\pm\pi/2$, and this resembles the quantum wire model proposed by Kitaev \cite{27}. 

Now consider the general case of a two-boundary system: the coin parameter is $\theta_1$ for $n<0$, and $\theta_2$ for sites from $n=0$ to $n=N$, and $\theta_3$ for $n>N$. We can now have two topologically distinct configurations. The symmetric configuration is defined by $\theta_3=\theta_1$ as shown in Fig.2(a) and the antisymmetric configuration is defined by $\theta_3=-\theta_1$, as shown in Fig.2(b). We can reduce this general case to the finite wire by letting $\theta_1$ approaches $\pm\pi/2$. 
Let's now use the following ansatz for the bound state wave function,
\begin{equation}
\begin{aligned}
\psi(n)&=Ae^{-\kappa_2 n}\left[\begin{array}{c}
a_{i\kappa_2}\\
b_{i\kappa_2}\end{array} \right] +Be^{\kappa_2 n}\left[\begin{array}{c}
a_{-i\kappa_2}\\
b_{-i\kappa_2}\end{array} \right]&0\leq n \leq N\\
&=Ce^{-\kappa_1 n}\left[\begin{array}{c}
\pm a_{i\kappa_1}\\
b_{i\kappa_1}\end{array} \right]&n>N\\
&=De^{\kappa_1 n}\left[\begin{array}{c}
a_{-i\kappa_1}\\
b_{-i\kappa_1}\end{array} \right]&n<0\\
\end{aligned}\label{eq:23}
\end{equation}
where $\kappa_1$ is positive so that the wave function vanishes at the infinity. The plus sign of the $\pm$ sign in the ansatz for $n>N$ is responsible for the symmetric configuration (Fig.2a), while minus sign is responsible for the antisymmetric one (Fig.2b). Here, the decay constant and the energy are related by Eq. \ref{eq:12}. The boundary conditions at $n=0$ and $n=N$ are 
\begin{equation}
\begin{aligned}
A b_{i\kappa_2}+Bb_{-i\kappa_2}&=Db_{-i\kappa_1}\\
Ae^{\kappa_2}a_{i\kappa_2}+Be^{-\kappa_2}a_{-i\kappa_2}&=De^{-\kappa_1}a_{-i\kappa_1}\\
A e^{-\kappa_2 (N+1)}b_{i\kappa_2}+B e^{\kappa_2 (N+1)}b_{-i\kappa_2}&=C e^{-\kappa_1 (N+1)}b_{i\kappa_1}\\
A e^{-\kappa_2 N}a_{i\kappa_2}+B e^{\kappa_2 N}a_{-i\kappa_2}&=\pm C e^{-\kappa_1 N}a_{i\kappa_1}
\end{aligned}\label{eq:24}
\end{equation}
This system of linear equations has nontrivial solution only when its determinant is zero. Thus, we obtain a relation between energy and decay constants. For the symmetric configuration (Fig.2a), this relation is,
\begin{equation}
\begin{aligned}
&\sinh{\left[\kappa_2 (N+1)\right]}(\sin^2{E}-\sin\theta_1\sin\theta_2)\\
=&\cosh{\left[\kappa_2 (N+1)\right]}\cos\theta_1\cos\theta_2\sinh\kappa_1\sinh\kappa_2
\end{aligned}\label{eq:25}
\end{equation}
while for the antisymmetric configuration (Fig.2b), it is,\begin{equation}
\begin{aligned}
\sin E(\cos\theta_1\sinh \kappa_1\tanh{\kappa_2 (N+1)}+\cos\theta_2\sinh \kappa_2)=0
\end{aligned}\label{eq:26}
\end{equation}
Notice that Eq. \ref{eq:26} is satisfied when $\sin E=0$, i.e. $E=0,\pi$, however, Eq. \ref{eq:25} does not have solution when $E=0, \pi$. This difference can be understood by looking at the spatial profile of the topological number shown in Fig.2. For the symmetric configuration (Fig.2a), there are two jumps in the topological number profile and is of special interest. Our discussion on boundary states indicates the existence of bound states localized at the boundaries with discontinuity in topological number. For the symmetric configuration with finite $N$, the system has bound states only when $\sin\theta_1$ and $\sin\theta_2$ are of opposite sign so that there exist a discontinuity in topological number profile. For two-boundary system with symmetric boundary condition, such as the two ends of a wire, the two Majorana states at the two ends interact with each other and this interaction between them gives rise to a small energy difference away from $E=0,\pi$. We substitute Eq. \ref{eq:12} into Eq. \ref{eq:25} and obtain the energy deviations from $E=0,\pi$ of the bound states with different separation $N$ and coin parameters $\theta_2$ different from $\theta_1=\pi/2$ (see Appendix II of the Supplementary material). 
The small energy deviations decrease exponentially by 
$\exp(-\kappa_2N)$ with $\kappa_2 = -\log{\left|(1-\mathrm{sgn}(\theta_2)\sin\theta_2)/\cos\theta_2\right|}$. This deviation is the interaction energy of the two Majorana state located at the two boundaries.
\begin{figure}
\includegraphics[scale = 0.55]{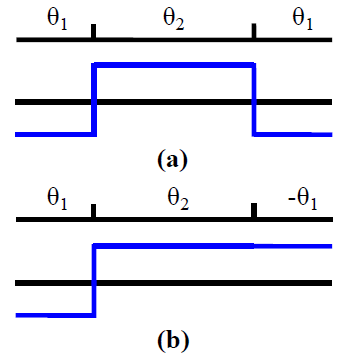}
\caption{\label{Figure 2}The spatial profile of the coin parameter (top) and topological number (bottom) for the symmetric (a) and antisymmetric (b) configuration. The topological profiles for the symmetric and antisymmetric configuration have one and two jumps respectively.}
\end{figure}
For the antisymmetric configuration (Fig.2b), there is only one jump in the topological number profile and thus, it is equivalent to the topological number profile of Majorana bound state obtained with one boundary. In this case, there are always bound state solutions with energy $E=0,\pi$. For example, for energy $E=0$ when $\theta_1<0$ and $\theta_2>0$, the bound state in the antisymmetric configuration is,
\begin{equation}
\begin{aligned}
&A=\sin\theta_1\quad\quad\quad\quad\quad\quad\quad B=0\\
&C=\sin\theta_2e^{(\kappa_1-\kappa_2)(N+1)}\quad D=\sin\theta_2\\
\end{aligned}
\end{equation}
This bound state solution has $B=0$ which means that the bound state is located at $n=0$ which is also the location of the jump in the topological number profile as shown in Fig. 2b. One can verify the solution with $E=\pi$ is also located at the discontinuous point of the topological number profile. In Fig. 3 , we illustrate the probability distribution of the bound states on a finite wire in three different configurations. The coin parameter on the line is $\theta_2 = -\pi/4$. For the symmetric configuration which coin parameters are $\pi/2$ on the two ends, the state is localized at the two ends. However, for the antisymmetric configuration which coin parameters are $\pi/2$ and $-\pi/2$ at the two ends, the state is localized at the end with coin parameter $\pi/2$.
\begin{figure}
\includegraphics[scale = 0.34]{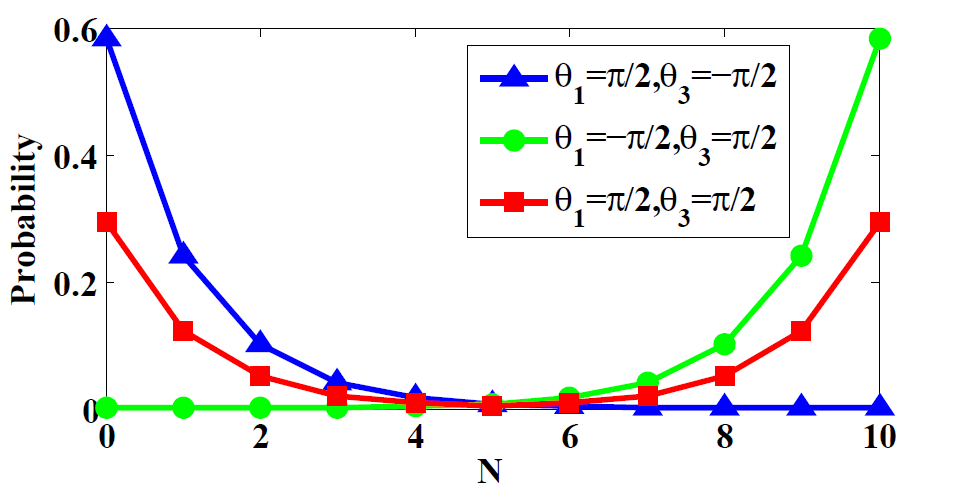}
\caption{\label{Figure 4}The probability distribution of boundary modes for three different configurations of a finite system with length $N=10$ and $\theta_2=-\pi/4$. The first and the second value in the legend is the coin parameter on the left and right end respectively.}
\end{figure}

- \textit{Conclusion} We have demonstrated the existence of two topological phases in a quantum walk system. We then investigated the boundary states for systems with single boundary in topological phases. Through an ansatz for bound state, we find solution for the boundary state and show that they are Majorana states with energies $E=0,\pi$. For systems with two boundaries in topological phases, such a  wire, we find same solution as the single boundary system if the boundary condition is antisymmetric. However, if the boundary condition is symmetric for the two boundary system, we find a small change in energy away from $E=0,\pi$, which is a signature of the interaction energy of the Majorana states. For the special case of a finite wire, the energies of the Majorana states have also been calculated as a function of the length of the wire. 
Possible observation of these bound states can be made following the work of Kitagawa et al.\cite{28} for photonic quantum walk. Our work may shed some insights into the recent experimental work on the interaction effects on a Majorana zero mode leaking into a quantum dot \cite{29,30}.  Experiments along these systems may lead to the observation of Majorana modes or boundary modes localized at the end of a finite quantum walk system either using photon or electrons, thereby verifying our theoretical calculation. On theoretical front, we like to seek analogy in a generalization of topological phases in higher-dimensional quantum walk system, as well as using these results for quantum computation. 
\begin{acknowledgments}
H. T. Lam would like to thank Hongliang Jiang, Fanqi Yuan and Kam Tuen Law for useful discussions. This work has been supported partially by  grant FSGRF13SC25 and FSGRF14SC28.
\end{acknowledgments}

\appendix*
\section{Supplementary Material}
\subsection{\label{Appendix I} Appendix I. Condition for the existence of bound state in single-boundary system}
For the existence of the bound state in single boundary system, there must be nontrivial solution of $r$ and $t$ of Eq. \ref{eq:13}. Thus we set the determinant of these linear equation to zero and this yields a condition between the energy $E$ and the decay constants:
\begin{equation}
\begin{aligned}
i\sin E=\frac{\sin\theta_2\cos\theta_1\sinh\kappa_1+\sin\theta_1\cos\theta_2\sinh\kappa_2}{\sin\theta_1-\sin\theta_2}\label{eq:17}
\end{aligned}
\end{equation}
As the right hand side of the equation is real while the left hand side is imaginary, equality only hold when $\sin E=0$, implying $E=0$ or $\pi$. The decay constant $\kappa$ in this case is
\begin{equation}
\begin{aligned}
&E=0,\quad e^{\kappa}=\frac{1\pm\sin\theta}{\cos\theta}\\
&E=\pi,\quad e^{\kappa}=-\frac{1\pm\sin\theta}{\cos\theta}\label{eq:18}
\end{aligned}
\end{equation}
Since at infinity the wave function decays to zero, $\kappa_1$ and $\kappa_2$ both must be positive, so that $\exp(\kappa)$ pick up the plus sign of the $\pm$ sign for positive $\sin\theta$ and the minus for negative $\sin\theta$. Using these $\kappa$ for $E=0,\pi$ in Eq. \ref{eq:17}, we obtain an equality,
\begin{equation}
\begin{aligned}
0=\sin\theta_2\sin\theta_1\frac{\mathrm{sign}(\sin\theta_1)+\mathrm{sign}(\sin\theta_2)}{\sin\theta_1-\sin\theta_2}\label{eq:19}
\end{aligned}
\end{equation}
However, this equation holds only when $\sin\theta_1$ and $\sin\theta_2$ are of different sign. Thus, our hypothetical bound state exists whenever the topological phases on the two sides of the boundary are different. Furthermore, there are always two bound states solution with energy $E=0,\pi$, which comes in pair in agreement with the prediction from the bulk-boundary correspondence.

\subsection{\label{Appendix II} Appendix II Condition for the existence of bound state in two-boundary system in symmetric configuration}
For two-boundary system in the symmetric configuration in the limit when $N$ tends to infinity, we have 
$\sinh\kappa_2 N\approx\mathrm{sign}(\kappa_2)\cosh\kappa_2 N$ 
and Eq. \ref{eq:25} becomes,
\begin{equation}
\begin{aligned}
\frac{\sin^2{E}-\sin\theta_1\sin\theta_2}{\cos\theta_1\cos\theta_2}=\mathrm{sgn}(\kappa_2)\sinh\kappa_1\sinh\kappa_2
\end{aligned}\label{eq:28}
\end{equation}
Equality can hold only when $\sin\theta_1$ and $\sin\theta_2$ has opposite sign because $\kappa_1$ is required to be positive from the boundary condition at infinity. This agrees with the topological condition in single boundary system. Now, when $\sin\theta_1$ and $\sin\theta_2$ has opposite sign, we find 
two solutions with energy $E=0,\pi$ with,
\begin{equation}
\begin{aligned}
\sinh\kappa_1 &= \mathrm{sign}(\theta_1)\tan\theta_1 \\ 
\sinh\kappa_2 &= \tan\theta_2
\end{aligned} \label{eq:29}
\end{equation}
\begin{table}
\caption{\label{tab:1} The bound state energies $E(\theta_2)$ of finite systems with length $N$, whose coin parameters are $\theta_2$ in the middle and $\theta_1=-\pi/2$ at the two ends as shown in Fig.2a. The system has four bound states solution with energy $E$, $-E$, $\pi-E$ and $-(\pi-E)$. }
\begin{center}
\begin{tabular}{ccccc}
\hline
$\quad N\quad$&$\quad E(\pi/3)/\pi\quad$&$\quad E(\pi/4)/\pi \quad$&$\quad E(\pi/6)/\pi\quad$\\
\hline
1	&$2.13\times 10^{-2}$		&$4.68\times 10^{-2}$		&$8.04\times 10^{-2}$\\
2	&$5.69\times 10^{-3}$		&$1.89\times 10^{-2}$		&$4.31\times 10^{-2}$\\
3	&$1.52\times 10^{-3}$		&$7.77\times 10^{-3}$		&$2.41\times 10^{-2}$\\
4	&$4.08\times 10^{-4}$		&$3.21\times 10^{-3}$		&$1.37\times 10^{-2}$\\
5	&$1.09\times 10^{-4}$		&$1.33\times 10^{-3}$		&$7.89\times 10^{-3}$\\
6	&$2.93\times 10^{-5}$		&$5.52\times 10^{-4}$		&$4.54\times 10^{-3}$\\
7	&$7.85\times 10^{-6}$		&$2.29\times 10^{-4}$		&$2.62\times 10^{-3}$\\
8	&$2.10\times 10^{-6}$		&$9.47\times 10^{-5}$		&$1.51\times 10^{-3}$\\
9	&$5.64\times 10^{-7}$		&$3.92\times 10^{-5}$		&$8.73\times 10^{-4}$\\
10	&$1.51\times 10^{-7}$		&$1.62\times 10^{-5}$		&$5.04\times 10^{-4}$\\
\hline
\end{tabular}
\end{center}
\end{table}

Similarly, for finite $N$, bound state solution can exist only when $\sin\theta_1$ and $\sin\theta_2$ are of opposite sign. From the dispersion relation Eq. \ref{eq:12}, we see that $|\sin E|$ is less than both $|\sin\theta_1|$ and $|\sin\theta_2|$ for a bound state solution and thus $(\sin^2 E-\sin\theta_1\sin\theta_2)$ is negative if $\sin\theta_1\sin\theta_2>0$. Note that the finite condition at infinity requires $\sinh\kappa_1>0$ so that equality of Eq. \ref{eq:25} requires $(\sin^2 E-\sin\theta_1\sin\theta_2)\geq 0$. Thus, for finite wire bound state solutions can exist only if $\sin\theta_1\sin\theta_2<0$. Using Eq. \ref{eq:12}, Eq. \ref{eq:25} can be written as an equation independent of $\kappa_1$,
\begin{equation}
\begin{aligned}
&\sinh{\left[\kappa_2 (N+1)\right]}(\sin^2{E}-\sin\theta_1\sin\theta_2)\\
=&\cosh{\left[\kappa_2 (N+1)\right]}\cos\theta_2\sinh\kappa_2\sqrt{\cos^2 E-\cos^2\theta_1}
\end{aligned}
\end{equation}
This new equation can be also applied to finite system where $\theta_1=\pi/2$ and $\kappa_1=+\infty$. The equation can now be solved by using Eq. \ref{eq:12} to relate $\kappa_1$ and $E$.
\end{document}